\documentclass[reprint,twocolumn]{revtex4}
\usepackage{amsfonts}
\usepackage{amsmath}
\usepackage{amssymb}
\usepackage{charter}
\usepackage{graphicx}
\usepackage{float}

\input{tcilatex}
\begin{document}

\title{Evaluating the quantum Ziv--Zakai bound in noisy environments}
\author{Shoukang Chang$^{1}$, Wei Ye$^{2}$, Xuan Rao$^{2}$, Huan Zhang$^{3}$%
, Mengmeng Luo$^{1}$, Yuetao Chen$^{1}$, Shaoyan Gao$^{1,\ast }$ and Liyun Hu%
$^{4,\dag }$}
\affiliation{$^{{\small 1}}$MOE Key Laboratory for Nonequilibrium Synthesis and
Modulation of Condensed Matter, Shaanxi Province Key Laboratory of Quantum
Information and Quantum Optoelectronic Devices, School of Physics, Xi'an
Jiaotong University, 710049, People's Republic of China\\
$^{{\small 2}}$School of Information Engineering, Nanchang Hangkong
University, Nanchang 330063, China\\
$^{{\small 3}}$School of Physics, Sun Yat-sen University, Guangzhou 510275,
China\\
$^{{\small 4}}$Center for Quantum Science and Technology, Jiangxi Normal
University, Nanchang 330022, China\\
$^{*,\dag }$Corresponding authors: gaosy@xjtu.edu.cn and hlyun2008@126.com}

\begin{abstract}
In the highly non-Gaussian regime, the quantum Ziv-Zakai bound (QZZB)
provides a lower bound on the available precision, demonstrating the better
performance compared with the quantum Cram\'{e}r-Rao bound. However,
evaluating the impact of a noisy environment on the QZZB without applying
certain approximations proposed by Tsang [Phys. Rev. Lett. 108, 230401
(2012)] remains a difficult challenge. In this paper, we not only derive the
general form of the QZZB with the photon loss and the phase diffusion by
invoking the technique of integration within an ordered product of
operators, but also show its estimation performance for several different
Gaussian resources, such as a coherent state (CS), a single-mode squeezed
vacuum state (SMSVS) and a two-mode squeezed vacuum state (TMSVS). Our
results indicate that compared with the SMSVS and the TMSVS, the QZZB for
the CS always shows the better estimation performance under the photon-loss
environment. More interestingly, for the phase-diffusion environment, the
estimation performance of the QZZB for the TMSVS can be better than that for
the CS throughout a wide range of phase-diffusion strength. Our findings
will provide a useful guidance for investigating the noisy quantum parameter
estimation.

\end{abstract}

\maketitle

\section{Introduction}

Fundamental principles of quantum mechanics, e.g., Heisenberg uncertainty,
impose ultimate precision limits on the parameter estimation \cite{1,2,3,4}.
To efficiently quantify the minimum estimation error in quantum metrology,
the quantum Cram\'{e}r-Rao bound (QCRB) is particularly famous for giving a
method to derive the asymptotically attainable estimation precision \cite%
{1,5,6,7}. More specifically, the QCRB is inversely proportional to the
quantum Fisher information (QFI), so that such a lower bound plays more
important role in metrologic applications, such as quantum sensing \cite%
{8,9,10}, gravitational wave detection \cite{11,12} and optical imaging \cite%
{13,14,15}. Especially, with the help of the multiparameter QCRB
corresponding to the QFI matrix, T.J. Proctor et al. investigated the
multiparameter estimation in the framework of networked\ quantum sensors 
\cite{16}. However, the QCRB is asymptotically tight only under the limit of
infinitely many trials, which may seriously underestimate the estimation
precision if the likelihood function is highly non-Gaussian \cite%
{17,18,19,20,21}. Thus,\ it is still an open problem to solve the bound of
evaluation accuracy for the limited number of tests or non-Gaussian cases.

For this reason, the quantum Weiss-Weinstein bound \cite{18} and the QZZB 
\cite{17,19} are often viewed as an alternative candidate to solve the
problem mentioned above. Compared to the former, the later has been widely
studied because it is relatively easy to calculate. For the single-parameter
estimation, V. Giovannetti et al. demonstrated how to obtain a lower bound
from prior information regimes, indicating that the sub-Heisenberg
estimation strategies are ineffective \cite{19}.\ After that, by relying on
the QZZB, Y. Gao and H. Lee theoretically derived the generalized limits for
parameter sensitivity when considering the implementations of adaptive
measurements, and they found that the precision of phase estimation with
several known states cannot be superior to the Heisenberg limit \cite{21}.
By extending the QZZB into the multiparameter cases, Y. R. Zhang and H. Fan
presented two kinds of metrological lower bounds using two different
approximations proposed by Tsang even in noisy systems \cite{20}. They
showed the advantage of simultaneous estimation over optimal individual one,
but the achievable bounds may be not tight either due to some approximation
methods used. Additionally, in order to further develop the QZZB, D. W.
Berry et al. proposed a novel lower bound for phase waveform estimation as
an application of multiparameter case via the quantum Bell-Ziv--Zakai bound 
\cite{22}. These results show that the QZZB become one of the promising
candidates to attain the lower bound of the (multi-)parameter estimation for
tightness.

On the other hand, for realistic scenarioes, because of the inevitable
interactions between the quantum system and its surrounding noisy
environment, the corresponding estimation precision would be reduced, which
has been studied extensively in recent years \cite%
{23,24,25,26,27,28,29,30,31}. In particular, since a variational method was
first proposed by Escher \cite{24},\ the analytical QCRB of single-(or
multiple-) parameter estimation in noisy environment can be derived
effectively \cite{32,33,34}. We also noticed that the effects of noisy
environment on the\ single-estimation performance of the QZZB has not been
studied before. Thus, in this paper, we shall focus on the general
derivation of the QZZB in the presence of both the photon loss and the phase
diffusion with the help of the technique of integration within an ordered
product of operators (IWOP) \cite{35,36,37,38,39,40,41}. Further, we also
present the estimation performance of the QZZB in the presence of the two
noise scenarioes when given some optical resources, such as the CS, the
SMSVS and the TMSVS. The results show that for the the photon-loss scenario,
the CS shows the better estimation performance of the QZZB, due to its
robustness against the photon losses, when comparing to the cases of both
the SMSVS and the TMSVS. While for the phase-diffusion scenario, the
estimation performance of the QZZB for the CS can be outperformed by that
for the TMSVS at the large range of the phase-diffusion strength.

This paper is arranged as follows. In section II, we briefly review the
known results of the QZZB. Based on the variational method and the IWOP
technique, in section III and IV, we respectively derive the tight QZZB in
the presence of the photon-loss and phase-diffusion scenarioes, and then
also investigate the estimation performance of the QZZB with the two noise
scenarioes for given some Gaussian states, such as the CS, the SMSVS and the
TMSVS. Finally, the main conclusions are drawn in the last section.

\section{The QZZB}

It is pointed out that the QZZB can show much tighter than the conventional
QCRB for the highly non-Gaussian regime \cite{17,18,19}. So, in this
section, we briefly review the known results of the quantum parameter
estimation based on the QZZB. From the perspective of a classical
parameter-estimated theory, let $x$ be the unknown parameter to be
estimated, $y$ be the observation with finite measurements, and $X\left(
y\right) $ be an estimator of $x$ constructed from the observation $y$.
Thus, the parameter sensitivity of $x$ can be quantified using the weighted
root mean square error 
\begin{equation}
\sum =\int dxdyp\left( y|x\right) p(x)[X\left( y\right) -x]^{2},  \label{1}
\end{equation}%
where $p\left( y|x\right) $ represents the condition probability density of
achieving the observation $y$ given $x$, and $p(x)$ represents the prior
probability density. According to Refs. \cite{42,43,44,45}, a classical
Ziv-Zakai bound, i.e., a lower bound for $\sum $, can be given by%
\begin{eqnarray}
\sum &\geq &\int_{0}^{\infty }d\beta \frac{\beta }{2}\chi \int_{-\infty
}^{\infty }dx2\min [p(x),p(x+\beta )]  \notag \\
&&\times \Pr \nolimits_{e}^{el}(x,x+\beta ),  \label{2}
\end{eqnarray}%
where $\Pr \nolimits_{e}^{el}(x,x+\beta )$ represents the minimum error
probability for the binary decision problem with an equally prior
probability, and $\chi $ is the so-called optional valley-filling operation,
i.e., $\chi f(\beta )\equiv \max_{\eta \geq 0}f(\beta +\eta )$, which
renders the bound tighter but harder to calculate \cite{17,42,43}.

Likewise, for the quantum parameter-estimated theory, let $\hat{\rho}\left(
x\right) $ be the density operator as a function of the unknown parameter $x$%
, and let $\hat{E}(y)$ be the positive operator-valued measure so as to
establish the measurement model. Then, the observation density is denoted as

\begin{equation}
p\left( y|x\right) =\text{Tr}[\hat{\rho}\left( x\right) \hat{E}(y)],
\label{3}
\end{equation}%
with a symbol of Tr being the operator trace. Thus, according to Refs. \cite%
{17,20,46}, a lower bound of the minimum error probability $\Pr
\nolimits_{e}^{el}(x,x+\beta )$ is given by

\begin{eqnarray}
\Pr \nolimits_{e}^{el}(x,x+\beta ) &\geq &\frac{1}{2}(1-\frac{1}{2}\left
\Vert \hat{\rho}\left( x\right) -\hat{\rho}\left( x+\beta \right) \right
\Vert _{1})  \notag \\
&\geq &\frac{1}{2}[1-\sqrt{1-F(\hat{\rho}\left( x\right) ,\hat{\rho}\left(
x+\beta \right) )}],  \label{4}
\end{eqnarray}%
with the trace norm $\left \Vert \hat{O}\right \Vert _{1}=$Tr$\sqrt{\hat{O}%
^{\dagger }\hat{O}}$ and the Uhlmann fidelity $F(\hat{\rho}\left( x\right) ,%
\hat{\rho}\left( x+\beta \right) )=(Tr\sqrt{\sqrt{\hat{\rho}\left( x\right) }%
\hat{\rho}\left( x+\beta \right) \sqrt{\hat{\rho}\left( x\right) }})^{2}$.

Now, let us assume that an unknown parameter $x$ is encoded into the quantum
state $\hat{\rho}\left( x\right) $, which can be presented in terms of an
unitary evolution%
\begin{equation}
\hat{\rho}\left( x\right) =e^{-i\hat{H}x}\hat{\rho}e^{i\hat{H}x},  \label{5}
\end{equation}%
where $\hat{\rho}$ is the initial state and $\hat{H}$ is the effective
Hamiltonian operator. Thus, it can be seen that $F(\hat{\rho}\left( x\right)
,\hat{\rho}\left( x+\beta \right) )\geq \left \vert \text{Tr}(\hat{\rho}e^{-i%
\hat{H}\beta })\right \vert ^{2}$ \cite{47}$.$ After assuming that the prior
probability density $p(x)$ is the uniform window with the mean $\mu $ and
the width $W$ denoted as%
\begin{equation}
p(x)=\frac{1}{W}rect\left( \frac{x-\mu }{W}\right) ,  \label{6}
\end{equation}%
and omitting the optional $\chi $, one can obtain the QZZB, i.e. \cite{17},

\begin{equation}
\sum \geq \sum \nolimits_{Z}=\int_{0}^{W}d\beta \frac{\beta }{2}\left( 1-%
\frac{\beta }{W}\right) [1-\sqrt{1-F(\beta )}],  \label{7}
\end{equation}%
where $F(\beta )\equiv \left \vert \text{Tr}(\hat{\rho}e^{-i\hat{H}\beta
})\right \vert ^{2}$ is a lower bound of the fidelity $F(\hat{\rho}\left(
x\right) ,\hat{\rho}\left( x+\beta \right) ).$ For the initial pure state $%
\hat{\rho},$ $F(\beta )$ is the fidelity between $\hat{\rho}\left( x\right) $
and $\hat{\rho}\left( x+\beta \right) .$ In the following, we would take the
effects of the environment on the QZZB into account, since the encoding
process of the quantum state $\hat{\rho}$ to unknown parameter $x$ is
inevitably affected by the environment, such as the photon losses and the
phase diffusion.

\section{The effects of photon losses on the QZZB}

In the case of photon losses, the encoding process of the quantum state to
unknown parameter $x$ no longer satisfies unitary evolution, so that the
QZZB can not be directly derived by using Eq. (\ref{7}). Fortunately,
similar to obtain the upper bound for QFI proposed by Escher with the
assistance of an variational method \cite{24,48}, combining with the IWOP
technique, here we shall present the derivation of the QZZB in the presence
of photon losses.

In order to change the encoding process into the unitary evolution $\hat{U}%
_{S+E}\left( x\right) $, the basic idea is to introduce additional degrees
of freedom, acting as an environment $E$ for the system $S$. When given an
initial pure state $\hat{\rho}_{S}=\left \vert \psi _{S}\right \rangle
\left
\langle \psi _{S}\right \vert $ of a probe system $S,$ the encoding
process of the initial pure state $\hat{\rho}_{S}$ is the non-unitary
evolution under the photon losses. So, it is necessary to expand the size of
the Hilbert space $S$ together with the photon losses environment space $E$.
After the quantum state in the enlarged space $S+E$ goes through the unitary
evolution $\hat{U}_{S+E}\left( x\right) ,$ one can obtain

\begin{eqnarray}
\hat{\rho}_{S+E}\left( x\right) &=&\left \vert \psi _{S+E}\left( x\right)
\right \rangle \left \langle \psi _{S+E}\left( x\right) \right \vert  \notag
\\
&=&\hat{U}_{S+E}\left( x\right) \hat{\rho}_{S}\otimes \hat{\rho}_{E_{0}}\hat{%
U}_{S+E}^{\dagger }\left( x\right)  \notag \\
&=&\sum_{l=0}^{\infty }\hat{\Pi}_{l}\left( x\right) \hat{\rho}_{S}\otimes 
\hat{\rho}_{E_{l}}\hat{\Pi}_{l}^{\dagger }\left( x\right) ,  \label{8}
\end{eqnarray}%
where $\hat{\rho}_{E_{0}}=\left \vert 0_{E}\right \rangle \left \langle
0_{E}\right \vert $ is the initial state of the photon losses environment
space $E$, $\hat{\rho}_{E_{l}}=\left \vert l_{E}\right \rangle \left \langle
l_{E}\right \vert $ is the orthogonal basis of the $\hat{\rho}_{E_{0}},$ and 
$\hat{\Pi}_{l}\left( x\right) $ is the Kraus operator acting on the $\hat{%
\rho}_{S},$ which can be described as

\begin{equation}
\hat{\Pi}_{l}\left( x\right) =\sqrt{\frac{\left( 1-\eta \right) ^{l}}{l!}}%
e^{-ix\left( \hat{n}-\lambda _{1}l\right) }\eta ^{\frac{\hat{n}}{2}}\hat{a}%
^{^{l}},  \label{9}
\end{equation}%
with the strength of photon losses $\eta $, the variational parameter $%
\lambda _{1},$ and the photon number operator $\hat{n}=\hat{a}^{\dag }\hat{a}
$. Note that $\eta =0$ and $\eta =1$ respectively correspond to complete
absorption and non-loss case. According to the Uhlmann's theorem \cite{49},
thus, the fidelity for the enlarged system $S+E$ with photon losses can be
given by%
\begin{equation}
F_{L_{1}}(\beta )=\max_{\left \{ \hat{\Pi}_{l}\left( x\right) \right \}
}F_{Q_{1}}(\hat{\rho}_{S+E}\left( x\right) ,\hat{\rho}_{S+E}\left( x+\beta
\right) ),  \label{10}
\end{equation}%
where 
\begin{eqnarray}
&&F_{Q_{1}}(\hat{\rho}_{S+E}\left( x\right) ,\hat{\rho}_{S+E}\left( x+\beta
\right) )  \notag \\
&=&\left \vert \left \langle \psi _{S+E}\left( x\right) |\psi _{S+E}\left(
x+\beta \right) \right \rangle \right \vert ^{2}  \notag \\
&=&\left \vert \text{Tr}(\hat{\rho}_{S}\hat{Z})\right \vert ^{2},  \label{11}
\end{eqnarray}%
corresponds to the lower bound of the fidelity in photon losses with $\hat{Z}
$ defined as \ 
\begin{equation}
\hat{Z}=\sum_{l=0}^{\infty }\hat{\Pi}_{l}^{\dagger }\left( x\right) \hat{\Pi}%
_{l}\left( x+\beta \right) .  \label{12}
\end{equation}%
By invoking the IWOP technique and Eqs. (\ref{9}) and (\ref{12}), one can
respectively obtain the operator identities, i.e.,

\begin{equation}
\eta ^{\hat{n}}e^{-i\beta \hat{n}}=\colon \exp \left[ \left( \eta e^{-i\beta
}-1\right) \hat{n}\right] \colon ,  \label{13}
\end{equation}

\begin{equation}
\hat{Z}=\left[ \eta e^{-i\beta }+\left( 1-\eta \right) e^{i\beta \lambda
_{1}}\right] ^{\hat{n}},  \label{14}
\end{equation}%
where $:\cdot :$ denotes the symbol of the normal ordering form. The more
details for the derivation can be seen in the Appendix A. Thus, based on
Eqs. (\ref{13}) and (\ref{14}), the lower bound of the fidelity, i.e., $%
F_{Q_{1}}(\hat{\rho}_{S+E}\left( x\right) ,\hat{\rho}_{S+E}\left( x+\beta
\right) )\equiv F_{Q_{1}}\left( \beta \right) ,$ can be given by

\begin{equation}
F_{Q_{1}}\left( \beta \right) =\left \vert \left \langle \psi _{S}\right
\vert \left[ \eta e^{-i\beta }+\left( 1-\eta \right) e^{i\beta \lambda _{1}}%
\right] ^{\hat{n}}\left \vert \psi _{S}\right \rangle \right \vert ^{2}.
\label{15}
\end{equation}%
From Eq. (\ref{15}), it should be noted that, when the variational parameter 
$\lambda _{1}$ takes the optimal value $\lambda _{1opt}$, $F_{Q_{1}}\left(
\beta \right) $ can\ reach the maximum value, which is the fidelity (denoted
as $F_{L_{1}}(\beta )$) in the photon-loss environment. In this situation,
the lower limit of the minimum error probability $\Pr%
\nolimits_{e_{L_{1}}}^{el}(x,x+\beta )$ under the photon losses can be
expressed as

\begin{eqnarray}
\Pr \nolimits_{e_{L_{1}}}^{el}(x,x+\beta ) &\geq &\frac{1}{2}(1-\frac{1}{2}%
\left \Vert \rho _{S+E}\left( x\right) +\rho _{S+E}\left( x+\beta \right)
\right \Vert _{1})  \notag \\
&\geq &\frac{1}{2}[1-\sqrt{1-F_{L_{1}}(\beta )}].  \label{16}
\end{eqnarray}%
According to Eq. (\ref{6}), finally, we can obtain the QZZB in the presence
of the photon-loss environment

\begin{eqnarray}
\sum \nolimits_{L_{1}} &\geq &\sum \nolimits_{Z_{L_{1}}}=\int_{0}^{W}d\beta 
\frac{\beta }{2}\left( 1-\frac{\beta }{W}\right)  \notag \\
&&\lbrack 1-\sqrt{1-F_{L_{1}}(\beta )}].  \label{17}
\end{eqnarray}%
Futhermore, by utilizing the inequalities \cite{17} 
\begin{eqnarray}
1-\sqrt{1-F_{L_{1}}(\beta )} &\geq &\frac{F_{L_{1}}(\beta )}{2},  \notag \\
\beta \left( 1-\frac{\beta }{W}\right) &\geq &\frac{W}{4}\sin \frac{\pi
\beta }{W},  \label{18}
\end{eqnarray}%
with the fidelity $F_{L_{1}}(\beta )$ satisfying the conditions of $0\leq
F_{L_{1}}(\beta )\leq 1$ and $0\leq \beta \leq W,$ the Eq. (\ref{17}) can be
further rewritten as

\begin{equation}
\sum \nolimits_{Z_{L_{1}}}\geq \sum \nolimits_{Z_{L_{1}}}^{\prime }\text{=}%
\int_{0}^{W}\tilde{F}_{1}(\beta )d\beta ,  \label{19}
\end{equation}%
where $\tilde{F}_{1}(\beta )=\frac{W}{16}F_{L_{1}}(\beta )\sin (\pi \beta
/W) $ is denoted as the generalized fidelity under the photon-loss case.

In this situation, here we shall consider the QZZB under the photon losses
when inputting three\ initial states of the probe system $S$, involving the
CS (denoted as $\left \vert \psi _{S}\left( \alpha \right) \right \rangle $%
), the SMSVS (denoted as $\left \vert \psi _{S}\left( r_{1}\right)
\right
\rangle $) and the TMSVS (denoted as $\left \vert \psi _{S}\left(
r_{2}\right) \right \rangle $). Following the approach proposed by Tsang 
\cite{17}, and according to Eq. (\ref{19}), one can respectively derive the
QZZB of the given initial states in the presence of the photon-loss
environment, i.e. [see Appendix B], \ 

\begin{eqnarray}
\sum \nolimits_{Z_{L_{1}}(\alpha )} &\geq &\sum \nolimits_{Z_{L_{1}}(\alpha
)}^{\prime }\text{=}\frac{\pi ^{3/2}e^{-4\eta N_{\alpha }}}{8\sqrt{\eta
N_{\alpha }}}\text{erfi}(2\sqrt{\eta N_{\alpha }}),  \notag \\
\sum \nolimits_{Z_{L_{1}}(r_{1})} &\geq &\sum
\nolimits_{Z_{L_{1}}(r_{1})}^{\prime }\text{=}\int_{0}^{2\pi }\tilde{F}%
_{1}(\beta )_{(r_{1})}d\beta ,  \notag \\
\sum \nolimits_{Z_{L_{1}}(r_{2})} &\geq &\sum
\nolimits_{Z_{L_{1}}(r_{2})}^{\prime }\text{=}\int_{0}^{2\pi }\tilde{F}%
_{1}(\beta )_{(r_{2})}d\beta ,  \label{20}
\end{eqnarray}%
where $N_{\alpha }=\left \vert \alpha \right \vert ^{2}$ is the mean photon
number of the CS, erfi$(\epsilon )\equiv (2/\sqrt{\pi })\int_{0}^{\epsilon
}\exp (t^{2})dt,$ and the generalized fidelities of both the SMSVS and the
TMSVS are respectively given by 
\begin{eqnarray}
\tilde{F}_{1}(\beta )_{(r_{1})} &=&\frac{\pi }{8}F_{L_{1}}(\beta
)_{(r_{1})}\sin (\beta /2),  \notag \\
\tilde{F}_{1}(\beta )_{(r_{2})} &=&\frac{\pi }{8}F_{L_{1}}(\beta
)_{(r_{2})}\sin (\beta /2),  \label{21}
\end{eqnarray}%
with%
\begin{eqnarray}
F_{L_{1}}(\beta )_{(r_{1})} &=&\max_{\left \{ \lambda _{1}\right \}
}F_{Q_{1}}\left( \beta \right) _{(r_{1})},  \notag \\
F_{L_{1}}(\beta )_{(r_{2})} &=&\max_{\left \{ \lambda _{1}\right \}
}F_{Q_{1}}\left( \beta \right) _{(r_{2})}.  \label{22}
\end{eqnarray}%
Note that $F_{Q_{1}}\left( \beta \right) _{(r_{1})}$=$\left. 1\right/ \left
\vert \sqrt{1+N_{r_{1}}\left[ 1-\Upsilon ^{2}(\eta ,\beta ,\lambda _{1})%
\right] }\right \vert ^{2}$ with the mean photon number $N_{r_{1}}$=$\sinh
^{2}r_{1}$, and $F_{Q_{1}}\left( \beta \right) _{(r_{2})}$=$\left. 1\right/
\left \vert 1+\left. N_{r_{2}}\left[ 1-\Upsilon (\eta ,\beta ,\lambda _{1})%
\right] \right/ 2\right \vert ^{2}$ with the mean photon number $N_{r_{2}}$=$%
2\sinh ^{2}r_{2}$ are respectively the lower bound of the fidelity for the
SMSVS and the TMSVS, as well as $\Upsilon (\eta ,\beta ,\lambda _{1})$=$\eta
e^{-i\beta }+\left( 1-\eta \right) e^{i\beta \lambda _{1}}$. In particular,
according to Eq. (\ref{20}), when $\eta =1$ corresponding to the non-loss
case, the corresponding QZZB for the input CS is consistent with the
previous work \cite{17}.

In order to visually see the effects of photon losses on the QZZB, at a
fixed value of $N=5,$ we plot the QZZB $\sum $ as a function of the
photon-loss strength $\eta $\ for several different states, involving the CS
(black dashed line), the SMSVS (red dashed line), and the TMSVS (blue dashed
line), as shown in Fig. 1(a). The results show that, with the decrease of $%
\eta ,$ the value of the QZZB for the given states increases. Especially,
compared to the another states, the QZZB for the CS increases relatively
slowly, which means that the CS as the input is more conducive to reducing
the estimation uncertainty under the photon losses. Further, to evaluate the
gap between the ideal and photon-loss cases, at a fixed $\eta =0.5,$ we also
show the QZZB $\sum $ as a function of the mean photon number $N$ for
several input resources, i.e., the CS (black lines), the SMSVS (red lines),
and the TMSVS (blue lines), as pictured in Fig. 1(b). For comparison, the
solid lines correspond to the ideal cases. It is clearly seen that, for the
CS, the gap between ideal and photon-loss cases is the smallest, which
implies that the CS is more robust against photon losses than other input
resources at the same conditions. Moreover, compared with both the CS and
the TMSVS, the estimation performance of the QZZB for the SMSVS is the worst
under the ideal or photon-loss cases. The reasons for these phenomenons can
be explained in terms of the generalized fidelity $\tilde{F}_{1}(\beta )$
given in Eq. (\ref{19}). For this purpose, at fixed values of $N=5$ and $%
\eta =0.5,$ we consider the generalized fidelity $\tilde{F}_{1}(\beta )$ as
a function of $\beta $ for several different initial states, including the
CS (black lines), the SMSVS (red lines), and the TMSVS (blue lines), as
shown in Fig. 2. According to Eq. (\ref{19}), the area enclosed by the curve
lines and abscissa is the value of the QZZB, which implies that the larger
the area, the worse the estimation performance. Taking the ideal case as a
concrete example, we can clearly see that the area for the SMSVS\ (red
region) is the largest, followed by that for the TMSVS (blue region), and
then that for the CS (black region), which is also true for the photon
losses. 
\begin{figure}[tbp]
\label{Fig1} \centering \includegraphics[width=0.72\columnwidth]{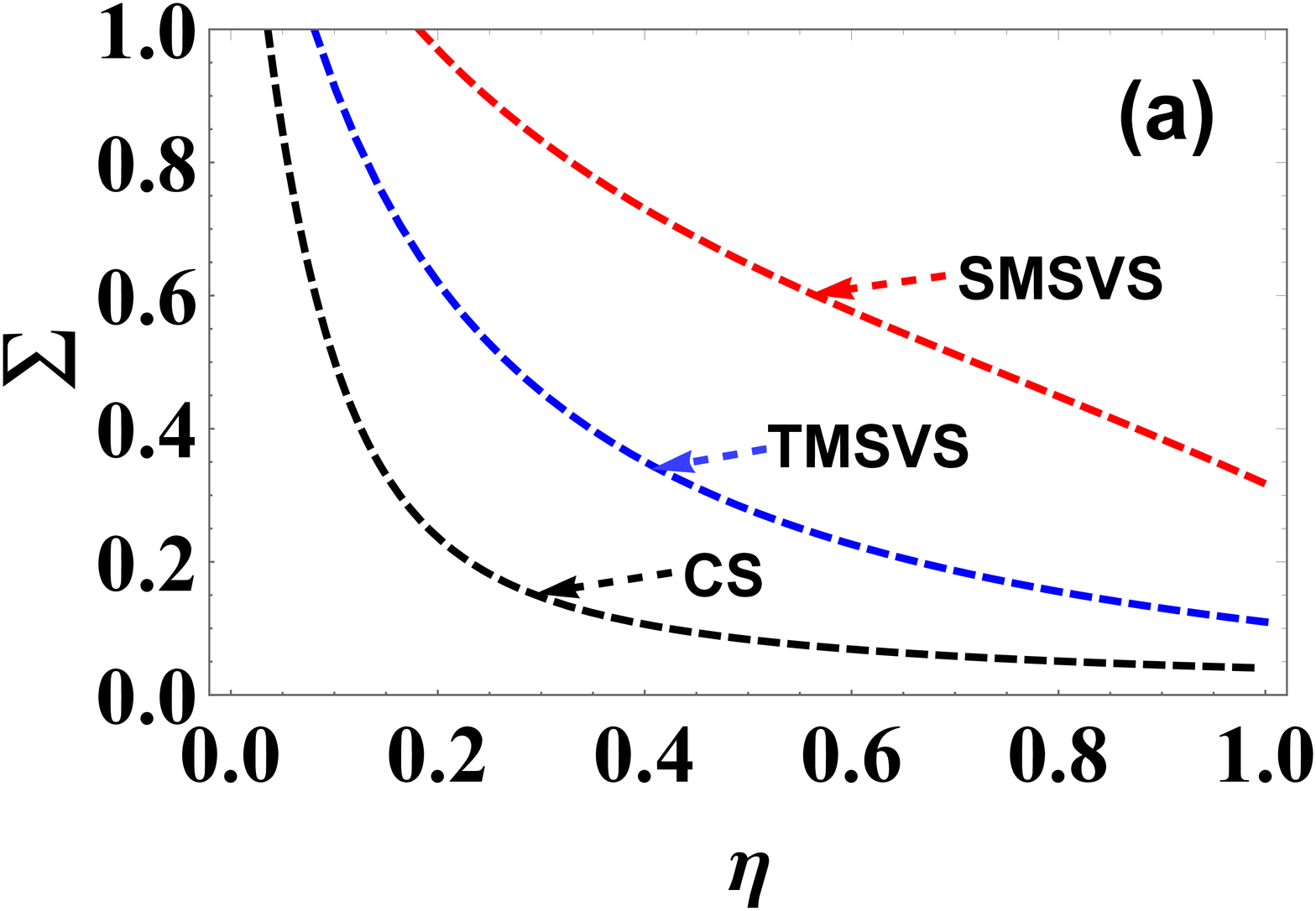} 
\newline
\includegraphics[width=0.73\columnwidth]{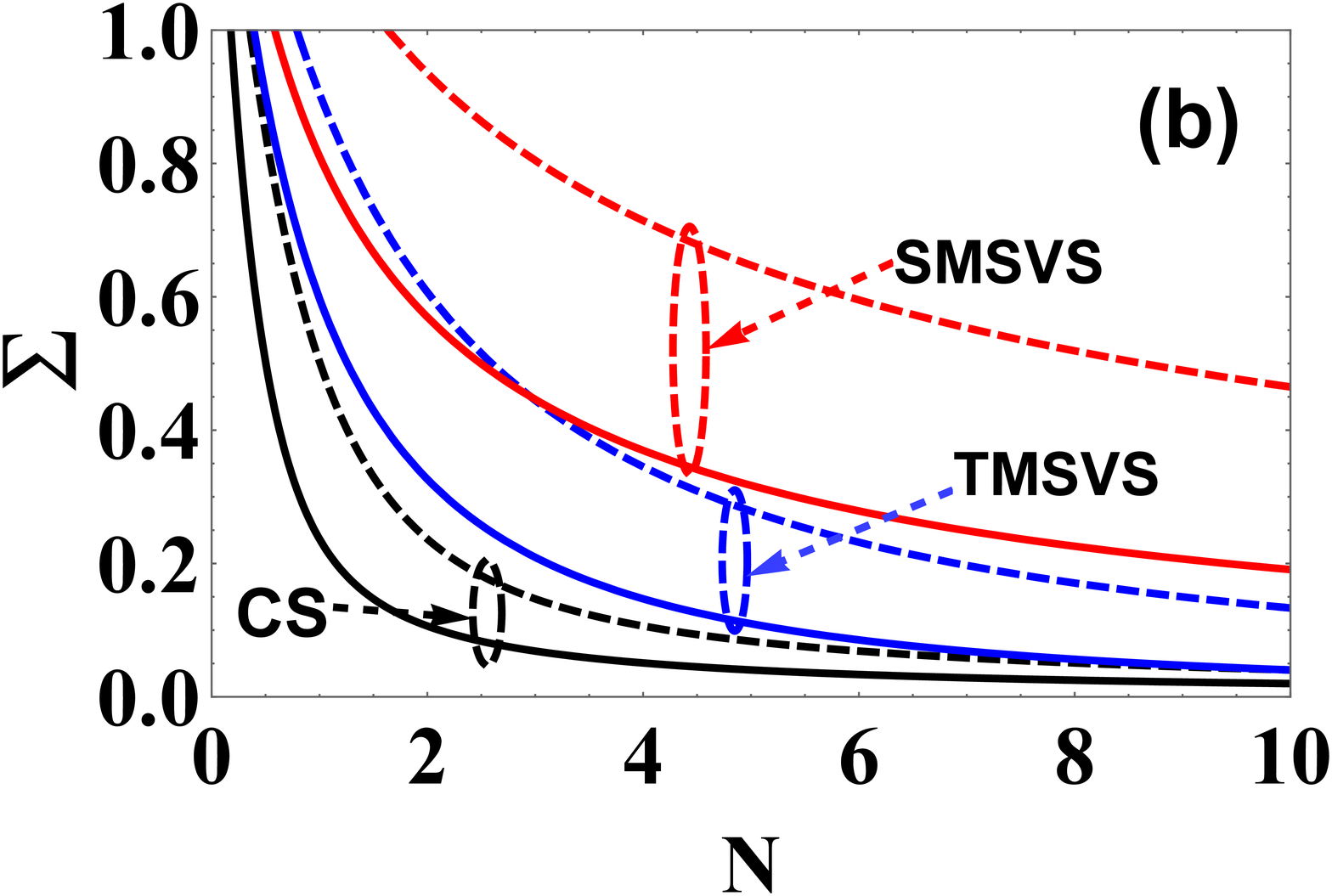} \newline
\caption{{}(Color online) The QZZB $\sum $ as a function of (a) the
photon-loss parameter $\protect \eta $ with $N=5$ and of (b) the mean photon
number $N$\ of initial state with $\protect \eta =0.5.$\ The black, blue and
red lines respectively correspond to CS, TMSVS and SMSVS as the initial
state. The dashed and solid lines correspond to the photon losses and no
photon losses, respectively. }
\end{figure}
\begin{figure}[tbp]
\label{Fig2} \centering \includegraphics[width=0.72\columnwidth]{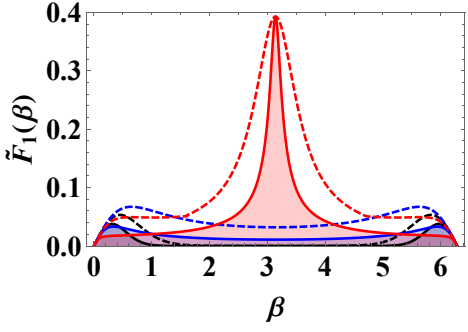} 
\newline
\caption{{}(Color online) The Generalized fidelity $\tilde{F}_{1}$ as a
function of the phase difference $\protect \beta $ with $N=5$ and $\protect%
\eta =0.5.$\ The black, blue and red lines respectively correspond to CS,
TMSVS and SMSVS as the initial state. The dashed and solid lines correspond
to the photon losses and no photon losses, respectively. }
\end{figure}

\section{The effects of phase diffusion on the QZZB}

More recently, the effects of phase diffusion on the QCRB have been studied
in Ref. \cite{48}, and they found that compared with photon losses, the
existence of phase diffusion has a greater influence on the QCRB. Naturally,
the question arises: what are the effects of phase diffusion on the
performance of the QZZB? To answer such a question, in this section, we
first derive the general form of the QZZB in the presence of phase diffusion
by using the variational method and the IWOP technique, and then show the
performance of the QZZB with the given states including the $\left \vert
\psi _{S}\left( \alpha \right) \right \rangle $, the $\left \vert \psi
_{S}\left( r_{1}\right) \right \rangle $ and the $\left \vert \psi
_{S}\left( r_{2}\right) \right \rangle $.

Generally speaking, the phase diffusion process can be modeled using the
interaction between the probe system $S$ and the environment $E^{\prime }$,
which can be described as%
\begin{equation}
\exp (i2\kappa \hat{n}\hat{q}_{E^{\prime }})=\exp [i\sqrt{2}\kappa \hat{n}(%
\hat{a}_{E^{\prime }}+\hat{a}_{E^{\prime }}^{\dagger })],  \label{23}
\end{equation}%
where $\kappa $ denotes the strength of phase diffusion and $\hat{q}%
_{E^{\prime }}=(\hat{a}_{E^{\prime }}+\hat{a}_{E^{\prime }}^{\dagger })/%
\sqrt{2}$ is the dimensionless position operator of the mirror. On this
background, the density operator $\hat{\rho}_{S+E^{\prime }}^{\prime }\left(
x\right) $ in the whole system $S+E^{\prime }$ can be given by%
\begin{eqnarray}
\hat{\rho}_{S+E^{\prime }}^{\prime }\left( x\right) &=&\left \vert \Phi
_{S+E^{\prime }}(x)\right \rangle \left \langle \Phi _{S+E^{\prime
}}(x)\right \vert  \notag \\
&=&\hat{U}_{S+E^{\prime }}^{\prime }\left( x\right) \hat{\rho}_{S}\otimes 
\hat{\rho}_{E_{0}^{\prime }}\hat{U}_{S+E^{\prime }}^{\prime \dagger }\left(
x\right)  \label{24}
\end{eqnarray}%
where $\hat{\rho}_{S}$ is the same as the aforementioned definition given in
Eq. (\ref{8}), $\hat{U}_{S+E^{\prime }}^{\prime }\left( x\right) =\exp
\left( -ix\hat{n}\right) \exp (i2\kappa \hat{n}\hat{q}_{E^{\prime }})$ is
the corresponding unitary operator of the combined system $S+E^{\prime },$
and $\hat{\rho}_{E_{0}^{\prime }}=\left \vert 0_{E^{\prime }}\right \rangle
\left \langle 0_{E^{\prime }}\right \vert $ is the initial state of the
phase-diffusion environment, which is often assumed to be the ground state
of a quantum oscillator. It is clearly seen from Eq. (\ref{24}) that the
density operator $\hat{\rho}_{S+E^{\prime }}^{\prime }\left( x\right) $ can
be viewed as the purifications of the probe state $\hat{\rho}_{S}\left(
x\right) $, which is given by \cite{48} 
\begin{eqnarray}
\hat{\rho}_{S}\left( x\right) &=&\text{Tr}_{E^{\prime }}\left[ \hat{\rho}%
_{S+E^{\prime }}^{\prime }\left( x\right) \right]  \notag \\
&=&\sum \limits_{m,n=0}^{\infty }\rho _{m,n}e^{-ix(m-n)-\kappa ^{2}(m-n)^{2}}
\notag \\
&&\times \left \vert m\right \rangle \left \langle n\right \vert ,
\label{25}
\end{eqnarray}%
where $\rho _{m,n}=\left \langle m\right \vert \hat{\rho}_{S}\left \vert
n\right \rangle $ is the matrix element of the initial state for the probe
system $S$. Thus, according to Eq. (\ref{24}), under the asymptotic
condition of $\sqrt{2}\kappa n\gg 1$, the purified unitary evolution can
integrally be written as \cite{48}

\begin{equation}
\hat{\rho}_{S+E^{\prime }}^{\prime \prime }\left( x\right) =\hat{u}%
_{E^{\prime }}(x)\hat{\rho}_{S+E^{\prime }}^{\prime }\left( x\right) \hat{u}%
_{E^{\prime }}^{\dagger }(x),  \label{26}
\end{equation}%
where $\hat{u}_{E^{\prime }}(x)=e^{\left. ix\lambda _{2}\hat{p}_{E^{\prime
}}\right/ 2\kappa }$ is the unitary operator with $\lambda _{2}$ being a
variational parameter and $\hat{p}_{E^{\prime }}=(\hat{a}_{E^{\prime }}-\hat{%
a}_{E^{\prime }}^{\dagger })/i\sqrt{2}$ being the dimensionless momentum
operator of the mirror, which acts only on the phase-diffusion environment $%
E^{\prime },$ and connects two purifications $\hat{\rho}_{S+E^{\prime
}}^{\prime }\left( x\right) $ and $\hat{\rho}_{S+E^{\prime }}^{\prime \prime
}\left( x\right) $ of the same probe state $\hat{\rho}_{S}\left( x\right) $.

Based on the Uhlmann's theorem \cite{49}, the fidelity in the
phase-diffusion environment $E^{\prime }$ can be given by

\begin{equation}
F_{L_{2}}(\beta )=\max_{\left \{ \left \vert \Phi _{S+E^{\prime }}(x)\right
\rangle \right \} }F_{Q_{2}}(\hat{\rho}_{S+E^{\prime }}^{\prime \prime
}\left( x\right) ,\hat{\rho}_{S+E^{\prime }}^{\prime \prime }\left( x+\beta
\right) ),  \label{27}
\end{equation}%
where $F_{Q_{2}}(\hat{\rho}_{S+E^{\prime }}^{\prime \prime }\left( x\right) ,%
\hat{\rho}_{S+E^{\prime }}^{\prime \prime }\left( x+\beta \right) )\equiv
F_{Q_{2}}\left( \beta \right) $ is the lower bound of the fidelity in the
phase-diffusion environment $E^{\prime },$ which can be derived as 
\begin{eqnarray}
F_{Q_{2}}\left( \beta \right) &=&\left \vert \left \langle \Psi
_{S+E^{\prime }}(x)|\Psi _{S+E^{\prime }}(x+\beta )\right \rangle \right
\vert ^{2}  \notag \\
&=&\Theta (\kappa ,\beta ,\lambda _{2})\left \vert \left \langle \psi
_{S}\right \vert e^{i\beta \left( \lambda _{2}-1\right) \hat{n}}\left \vert
\psi _{S}\right \rangle \right \vert ^{2},  \label{28}
\end{eqnarray}%
with $\Theta (\kappa ,\beta ,\lambda _{2})=e^{\left. -\beta ^{2}\lambda
_{2}^{2}\right/ 8\kappa ^{2}}$. One can refer to the Appendix C for more
details of the corresponding derivation in Eq. (\ref{28}).

Likewise, if the variational parameter $\lambda _{2}$ takes the optimal
value $\lambda _{2opt}$, then $F_{Q_{2}}\left( \beta \right) $ can achieve
the maximum value, which is the fidelity in the phase-diffusion environment $%
F_{L_{2}}(\beta ).$ Further, the lower limit of the minimum error
probability $\Pr \nolimits_{e_{L_{2}}}^{el}(x,x+\beta )$ in the presence of
the phase-diffusion environment $E^{\prime }$ can be given by%
\begin{eqnarray}
\Pr \nolimits_{e_{L_{2}}}^{el}(x,x+\beta ) &\geq &\frac{1}{2}(1-\frac{1}{2}%
\left \Vert \hat{\rho}_{S+E^{\prime }}^{\prime \prime }\left( x\right) +\hat{%
\rho}_{S+E^{\prime }}^{\prime \prime }\left( x+\beta \right) \right \Vert
_{1})  \notag \\
&\geq &\frac{1}{2}[1-\sqrt{1-F_{L_{2}}(\beta )}].  \label{29}
\end{eqnarray}%
Similar to how we get the Eq. (\ref{19}), the QZZB in the presence of the
phase-diffusion environment can be denoted as%
\begin{eqnarray}
\sum \nolimits_{L_{2}} &\geq &\sum \nolimits_{Z_{L_{2}}}=\int_{0}^{W}d\beta 
\frac{\beta }{2}\left( 1-\frac{\beta }{W}\right)  \notag \\
&&\lbrack 1-\sqrt{1-F_{L_{2}}(\beta )}].  \label{30}
\end{eqnarray}%
By using the Eq. (\ref{18}), finally, the Eq. (\ref{30}) can be rewritten as

\begin{equation}
\sum \nolimits_{Z_{L_{2}}}\geq \sum \nolimits_{Z_{L_{2}}}^{\prime
}=\int_{0}^{W}\tilde{F}_{2}(\beta )d\beta ,  \label{31}
\end{equation}%
where $\tilde{F}_{2}(\beta )=\frac{W}{16}F_{L_{2}}(\beta )\sin (\pi \beta
/W) $ is also denoted as the generalized fidelity under the phase-diffusion
case.

Next, let us consider the performance of the QZZB with the given states
including the $\left \vert \psi _{S}\left( \alpha \right) \right \rangle $,
the $\left \vert \psi _{S}\left( r_{1}\right) \right \rangle $ and the $%
\left \vert \psi _{S}\left( r_{2}\right) \right \rangle $. According to the
Eq. (\ref{31}), one can respectively derive the QZZB of these given states
in the presence of the phase-diffusion environment, i.e.,%
\begin{eqnarray}
\sum \nolimits_{Z_{L_{2}}\left( \alpha \right) } &\geq &\sum
\nolimits_{Z_{L_{2}}\left( \alpha \right) }^{\prime }\text{=}\int_{0}^{2\pi }%
\tilde{F}_{2}(\beta )_{\left( \alpha \right) }d\beta ,  \notag \\
\sum \nolimits_{Z_{L_{2}}\left( r_{1}\right) } &\geq &\sum
\nolimits_{Z_{L_{2}}\left( r_{1}\right) }^{\prime }\text{=}\int_{0}^{2\pi }%
\tilde{F}_{2}(\beta )_{\left( r_{1}\right) }d\beta ,  \notag \\
\sum \nolimits_{Z_{L_{2}}\left( r_{2}\right) } &\geq &\sum
\nolimits_{Z_{L_{2}}\left( r_{2}\right) }^{\prime }\text{=}\int_{0}^{2\pi }%
\tilde{F}_{2}(\beta )_{\left( r_{2}\right) }d\beta ,  \label{32}
\end{eqnarray}%
where we have set 
\begin{eqnarray}
\tilde{F}_{2}(\beta )_{\left( \alpha \right) } &\text{=}&\frac{\pi }{8}%
F_{L_{2}}(\beta )_{\left( \alpha \right) }\sin (\beta /2),  \notag \\
\tilde{F}_{2}(\beta )_{\left( r_{1}\right) } &\text{=}&\frac{\pi }{8}%
F_{L_{2}}(\beta )_{\left( r_{1}\right) }\sin (\beta /2),  \notag \\
\tilde{F}_{2}(\beta )_{\left( r_{2}\right) } &\text{=}&\frac{\pi }{8}%
F_{L_{2}}(\beta )_{\left( r_{2}\right) }\sin (\beta /2),  \label{33}
\end{eqnarray}%
with%
\begin{eqnarray}
F_{L_{2}}(\beta )_{\left( \alpha \right) } &\text{=}&\max_{\lambda
_{2}}\left \{ \Theta (\kappa ,\beta ,\lambda _{2})\exp \left[ -2N_{\alpha
}\Lambda (\beta ,\lambda _{2})\right] \right \} ,  \notag \\
F_{L_{2}}(\beta )_{\left( r_{1}\right) } &\text{=}&\max_{\lambda _{2}}\frac{%
\Theta (\kappa ,\beta ,\lambda _{2})}{\sqrt{1+2N_{r_{1}}\left(
1+N_{r_{1}}\right) \Lambda (2\beta ,\lambda _{2})}},  \notag \\
F_{L_{2}}(\beta )_{\left( r_{2}\right) } &\text{=}&\max_{\lambda _{2}}\frac{%
\Theta (\kappa ,\beta ,\lambda _{2})}{1+N_{r_{2}}\left( 1+N_{r_{2}}/2\right)
\Lambda (\beta ,\lambda _{2})},  \notag \\
\Lambda (\beta ,\lambda _{2}) &=&1-\cos \left[ \beta \left( \lambda
_{2}-1\right) \right] .  \label{34}
\end{eqnarray}

For the sake of clearly seeing the effects of phase diffusion on the QZZB,
at a fixed value of $N=5,$ we plot the QZZB $\sum $ as a function of the
phase-diffusion strength $\kappa $\ for the given states including the CS
(black dot-dashed line), the SMSVS (red dot-dashed line) and the TMSVS (blue
dot-dashed line), as shown in Fig. 3(a). It is clear that the value of the
QZZB for the given states increases with the increase of $\kappa $. In
particular, compared to other states, the corresponding QZZB for the SMSVS
is relatively larger and increases rapidly as the phase-diffusion strength $%
\kappa $ increases, which means that both the CS and the TMSVS instead of
the SMSVS is a better choice to the robustness against the phase diffusion.
More precisely, at the rang of $0\leqslant \kappa \leqslant 0.41,$ the value
of the QZZB for the CS can be lower than that for the TMSVS, but the former
can be larger than the latter when $\kappa $ is greater than $0.41$. This
phenomenon means that the CS can be more sensitive to the phase-diffusion
environment compared to the TMSVS [see Fig. 3(b)]. In addition, under the
phase diffusion processes (e.g., $\kappa =0.2$), we further consider the
QZZB $\sum $ as a function of the mean photon number $N$ for the CS (black
lines), the SMSVS (red lines), and the TMSVS (blue lines), as shown in Fig.
3(b). As a comparison, the ideal cases (solid lines) are also plotted here.
It is shown that, the gap with the SMSVS (red lines) between the ideal and
phase diffusion cases is the largest, which means that the SMSVS is more
sensitive to the phase diffusion than other states. We also find that the
gap with the TMSVS is smaller than the one with the CS when given the same
mean photon number $N$, which does exist in the case of photon losses. Even
so, the estimation performance of the QZZB for the CS is superior to that
for the TMSVS under the phase diffusion. Similar to Fig. 2, in order to
better explain these phenomena, at fixed values of $N=5$ and $\kappa =0.2,$
Fig. 4 shows the generalized fidelity $\tilde{F}_{2}(\beta )$ as a function
of $\beta $, in which the area enclosed by the curve lines and abscissa is
the value of the QZZB. Likewise, by taking the phase diffusion as an
example, the area for the SMSVS\ (red region) is the largest, followed by
that for the TMSVS (blue region), and then that for the CS (black region),
which implies that the CS shows the best estimation performance in the
presence of the phase diffusion. 
\begin{figure}[tbp]
\label{Fig3} \centering \includegraphics[width=0.72\columnwidth]{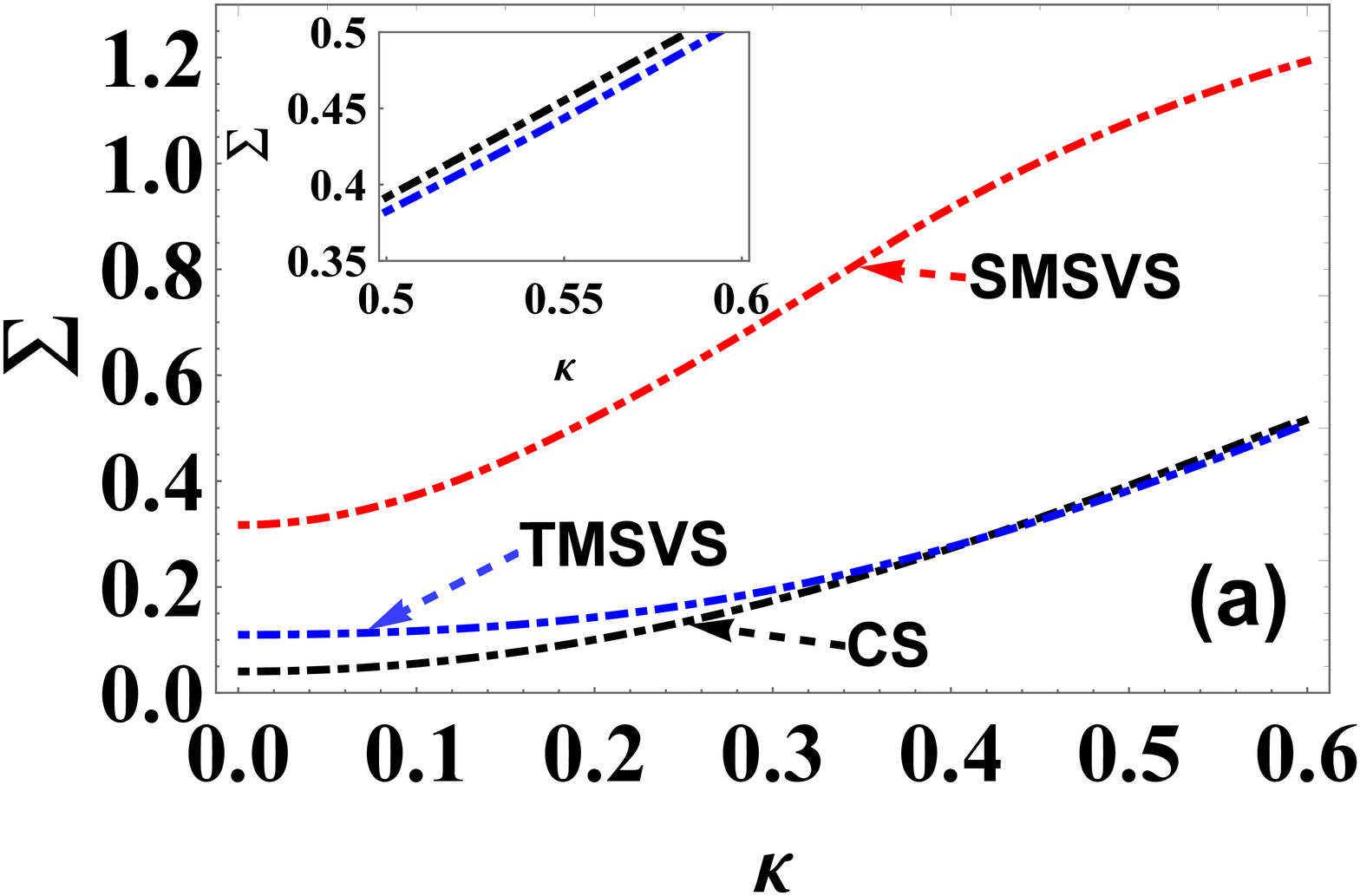} 
\newline
\includegraphics[width=0.73\columnwidth]{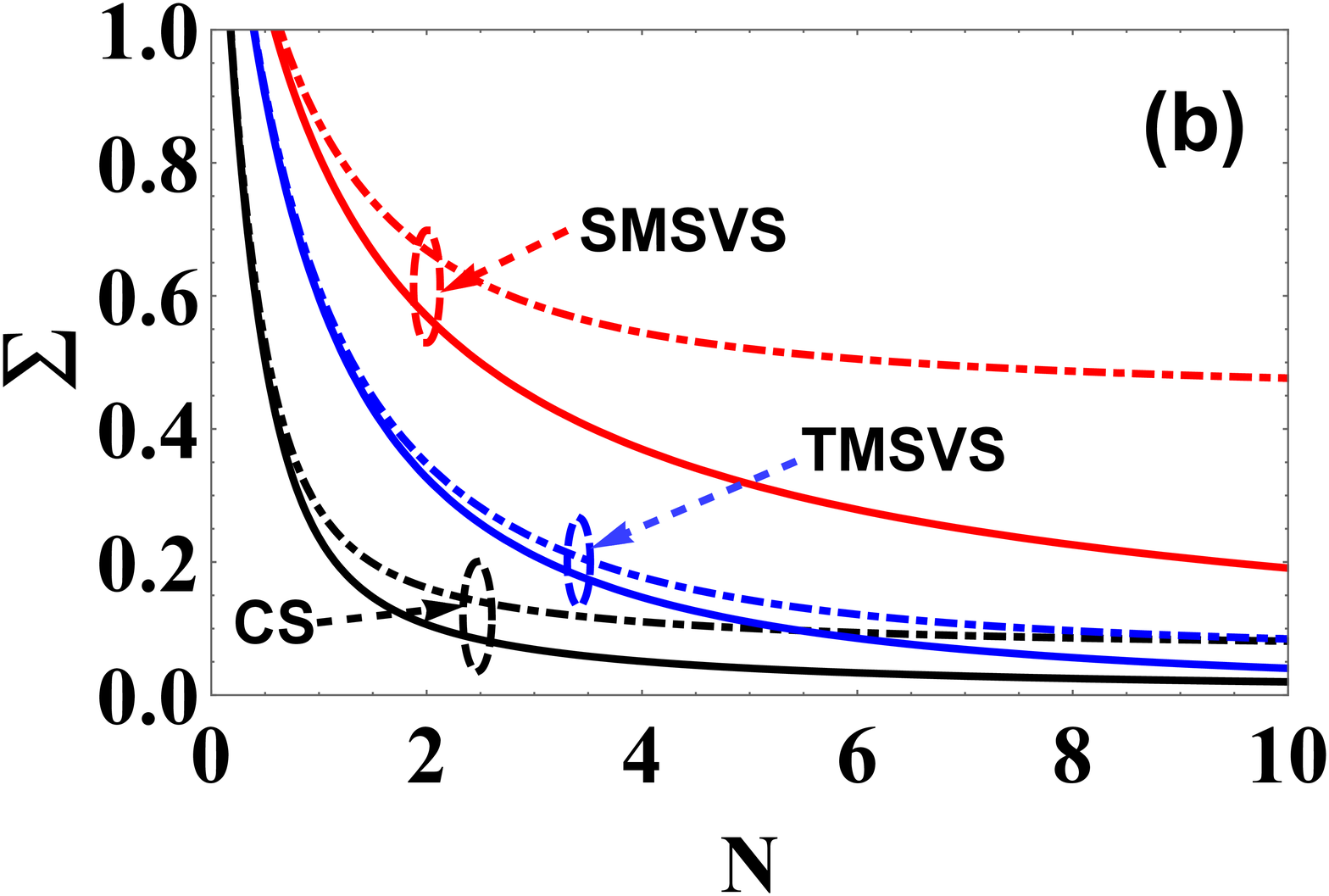} \newline
\caption{{}(Color online) The QZZB $\sum $ as a function of (a) the phase
diffusion parameter $\protect \kappa $ with $N=5$ and of (b) the mean photon
number $N$\ of initial state with $\protect \kappa =0.2.$\ The black, blue
and red lines respectively correspond to CS, TMSVS and SMSVS as the initial
state. The dot-dashed and solid lines correspond to the phase diffusion and
no phase diffusion, respectively. }
\end{figure}
\begin{figure}[tbp]
\label{Fig4} \centering \includegraphics[width=0.72\columnwidth]{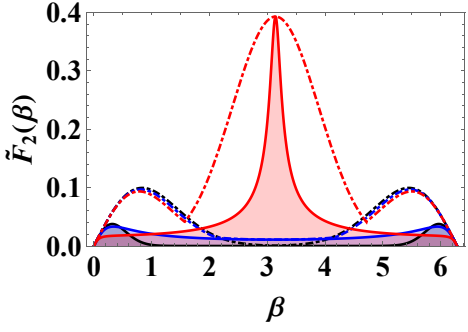} 
\newline
\newline
\caption{{}(Color online) The Generalized fidelity $\tilde{F}_{2}$ as a
function of the phase difference $\protect \beta $ with $N=5$ and $\protect%
\kappa =0.2.$ The black, blue and red lines respectively correspond to CS,
TMSVS and SMSVS as the initial state. The dot-dashed and solid lines
correspond to the phase diffusion and no phase diffusion, respectively. }
\end{figure}

\section{Conclusions}

In summary, we have presented the estimation performance of the QZZB in
noisy systems with several different Gaussian resources, including the CS,
the SMSV and the TMSV. One of the major contributions showed in this paper
is that the analytical expression of the QZZB is completely derived in the
presence of the photon losses. It is shown that the estimation performance
of the QZZB is related to the generalized fidelity, which can be used to
explain the phenomenon that the QZZB for both the CS and the TMSVS performs
better than that for the SMSVS, especially for the best performance with
inputting the CS. Furthermore, we also investigate the effects of the phase
diffusion systems on the QZZB with the same Gaussian resources. It is found
that the CS as the initial state can show the better estimation performance
of the QZZB at the small range of the phase-diffusion strength ($0\leqslant
\kappa \leqslant 0.41$), but for the large range of the phase-diffusion
strength $\left( \kappa >0.41\right) ,$ its estimation performance can be
exceeded by that for the TMSVS. Finally, we should mention that it is
impossible for investigating the optimal estimation performance by using the
QZZB \cite{20}, which is still a crucial problem for the future work.

\begin{acknowledgments}
This work was supported by National Nature Science Foundation of China
(Grant Nos. 91536115, 11534008, 11964013, 11664017, 62161029), Natural
Science Foundation of Shaanxi Province (Grant No. 2016JM1005), the Training
Program for Academic and Technical Leaders of Major Disciplines in Jiangxi
Province (Grant No. 20204BCJL22053), and Natural Science Foundation of
Jiangxi Provincial (Grant No. 20202BABL202002).
\end{acknowledgments}

\textbf{Appendix\ A: The Proof of Eqs. (\ref{13}) and (\ref{14})}

By using the completeness relation of Fock states, one can obtain%
\begin{align}
& \eta ^{\hat{n}}e^{-i\beta \hat{n}}  \notag \\
& =\eta ^{\hat{n}}e^{-i\beta \hat{n}}\sum_{k=0}^{\infty }\left \vert k\right
\rangle \left \langle k\right \vert  \notag \\
& =\sum_{k=0}^{\infty }\eta ^{k}e^{-i\beta k}\left \vert k\right \rangle
\left \langle k\right \vert  \notag \\
& =\sum_{k=0}^{\infty }\eta ^{k}e^{-i\beta k}\frac{(\hat{a}^{\dagger })^{k}}{%
k!}\left \vert 0\right \rangle \left \langle 0\right \vert \hat{a}^{k} 
\notag \\
& =\colon \sum_{k=0}^{\infty }\eta ^{k}e^{-i\beta k}\frac{(\hat{a}^{\dagger
})^{k}}{k!}e^{-\hat{a}^{\dagger }\hat{a}}\hat{a}^{k}\colon  \notag \\
& =\colon e^{-\hat{a}^{\dagger }\hat{a}}\sum_{k=0}^{\infty }\frac{(\eta
e^{-i\beta }\hat{a}^{\dagger }\hat{a})^{k}}{k!}\colon  \notag \\
& =\colon \exp \left[ \left( \eta e^{-i\beta }-1\right) \hat{n}\right] \colon
\tag{A1}
\end{align}%
where we have utilized the IWOP technique and the normal ordering form of
the vacuum projection operator \cite{36} 
\begin{equation}
\left \vert 0\right \rangle \left \langle 0\right \vert =\colon e^{-\hat{a}%
^{\dagger }\hat{a}}\colon .  \tag{A2}
\end{equation}%
Then, using Eq. (A1), one can further derive the operator $\hat{Z}$, i.e.,%
\begin{align}
\hat{Z}& =\sum_{l=0}^{\infty }\hat{\Pi}_{l}^{\dagger }\left( x\right) \hat{%
\Pi}_{l}\left( x+\beta \right)  \notag \\
& =\sum_{l=0}^{\infty }\frac{\left( 1-\eta \right) ^{l}}{l!}\left( \hat{a}%
^{\dagger }\right) ^{^{l}}\eta ^{\hat{n}}e^{-i\beta \left( \hat{n}-\lambda
_{1}l\right) }\hat{a}^{^{l}}  \notag \\
& =\sum_{l=0}^{\infty }\frac{[e^{i\beta \lambda _{1}}\left( 1-\eta \right)
]^{l}}{l!}\left( \hat{a}^{\dagger }\right) ^{^{l}}\eta ^{\hat{n}}e^{-i\beta 
\hat{n}}\hat{a}^{^{l}}  \notag \\
& =\colon \sum_{l=0}^{\infty }\frac{[e^{i\beta \lambda _{1}}\left( 1-\eta
\right) ]^{l}}{l!}\left( \hat{a}^{\dagger }\right) ^{^{l}}\exp \left[ \left(
\eta e^{-i\beta }-1\right) \hat{n}\right] \hat{a}^{^{l}}\colon  \notag \\
& =\colon \exp \left[ \left( \eta e^{-i\beta }-1\right) \hat{n}\right]
\sum_{l=0}^{\infty }\frac{[e^{i\beta \lambda _{1}}\left( 1-\eta \right) \hat{%
a}^{\dagger }\hat{a}]^{l}}{l!}\colon  \notag \\
& =\colon \exp \left[ \left( \eta e^{-i\beta }+\left( 1-\eta \right)
e^{i\beta \lambda _{1}}-1\right) \hat{n}\right] \colon  \notag \\
& =\left[ \eta e^{-i\beta }+\left( 1-\eta \right) e^{i\beta \lambda _{1}}%
\right] ^{\hat{n}}  \tag{A3}
\end{align}%
where we have utilized the following operator identity about $e^{\lambda 
\hat{a}^{\dagger }\hat{a}}$ , i.e., 
\begin{equation}
e^{\lambda \hat{a}^{\dagger }\hat{a}}=\colon e^{\left( e^{\lambda }-1\right) 
\hat{a}^{\dagger }\hat{a}}\colon ,  \tag{A4}
\end{equation}%
to remove the symbol of normal ordering.

\textbf{Appendix B: The} \textbf{QZZB for the CS in the presence of the
photon losses environment}

Based on Eq. (\ref{15}), one can get the lower bound of the fidelity for the
CS $\left \vert \psi _{S}\left( \alpha \right) \right \rangle $ under the
photon-loss environment%
\begin{align}
& F_{Q_{1}}\left( \beta \right) _{CS}  \notag \\
& =\exp \left[ 2N_{\alpha }\left( \eta \cos \beta +\left( 1-\eta \right)
\cos \beta \lambda _{1}-1\right) \right] .  \tag{B1}
\end{align}%
The value of the variational parameter $\lambda _{1}$ that maximizes the
lower bound is $\lambda _{1opt}=0,$ which yields the fidelity for the CS $%
\left \vert \psi _{S}\left( \alpha \right) \right \rangle $, i.e.,

\begin{equation}
F_{Q_{1}}\left( \beta \right) _{CS}=\exp \left[ 2\eta N_{\alpha }\left( \cos
\beta -1\right) \right] .  \tag{B2}
\end{equation}

Thus, substituting Eq. (B2) into Eq. (\ref{19}), and $W=2\pi $, one can get

\begin{align}
\sum \nolimits_{Z_{L_{1}}}& \geq \sum \nolimits_{Z_{L_{1}}}^{\prime }=\frac{%
\pi }{8}\int_{0}^{2\pi }\exp \left[ 2\eta N_{\alpha }\left( \cos \beta
-1\right) \right]  \notag \\
& \times \sin (\beta /2)d\beta ,  \tag{B3}
\end{align}%
Changing the integral variable $\beta $ to $s\equiv \cos (\beta /2)$ and
utilizing the identity $\cos \beta =2\cos ^{2}(\beta /2)-1,$ one can finally
obtain the QZZB for the CS in the presence of the photon-loss environment 
\begin{equation}
\sum \nolimits_{Z_{L_{1}}}\geq \sum \nolimits_{Z_{L_{1}}}^{\prime }=\frac{%
\pi ^{3/2}e^{-4\eta N_{\alpha }}}{8\sqrt{\eta N_{\alpha }}}\text{erfi}(2%
\sqrt{\eta N_{\alpha }}).  \tag{B4}
\end{equation}

Next, using the completeness relation of coherent states, the SMSVS $%
\left
\vert \psi _{S}\left( r_{1}\right) \right \rangle $ can be expanded
in the basis of the CS,

\begin{equation}
\left \vert \psi _{S}\left( r_{1}\right) \right \rangle =\sqrt{\sec hr_{1}}%
\int \frac{d^{2}z_{1}}{\pi }e^{-\frac{1}{2}\left \vert z_{1}\right \vert
^{2}+\frac{1}{2}z_{1}^{\ast 2}\tanh r_{1}}\left \vert z_{1}\right \rangle . 
\tag{B5}
\end{equation}%
Then, using Eq. (B5), one can obtain the lower bound of the fidelity for the
SMSVS in the presence of the photon-loss environment

\begin{align}
& F_{Q_{1}}\left( \beta \right) _{SMSV}  \notag \\
& =\left \vert \left \langle \psi _{S}\left( r_{1}\right) \right \vert \left[
\eta e^{-i\beta }+\left( 1-\eta \right) e^{i\beta \lambda _{1}}\right] ^{%
\hat{n}}\left \vert \psi _{S}\left( r_{1}\right) \right \rangle \right \vert
^{2}  \notag \\
& =|\sec hr_{1}\int \frac{d^{2}z_{1}d^{2}z_{2}}{\pi ^{2}}\exp [-\frac{1}{2}%
(\left \vert z_{1}\right \vert ^{2}+\left \vert z_{2}\right \vert ^{2}) 
\notag \\
& +\frac{1}{2}(z_{1}^{\ast 2}+z_{2}^{2})\tanh r_{1}]  \notag \\
& \times \left \langle z_{2}\right \vert \colon \exp \left[ \left( \eta
e^{-i\beta }+\left( 1-\eta \right) e^{i\beta \lambda _{1}}-1\right)
a^{\dagger }a\right] \colon  \notag \\
& \times \left \vert z_{1}\right \rangle |^{2}  \notag \\
& =|\sec hr_{1}\int \frac{d^{2}z_{1}d^{2}z_{2}}{\pi ^{2}}\exp [-\frac{1}{2}%
(\left \vert z_{1}\right \vert ^{2}+\left \vert z_{2}\right \vert ^{2}) 
\notag \\
& +\left( \eta e^{-i\beta }+\left( 1-\eta \right) e^{i\beta \lambda
_{1}}-1\right) z_{1}z_{2}^{\ast }  \notag \\
& +\frac{1}{2}(z_{1}^{\ast 2}+z_{2}^{2})\tanh r_{1}]|^{2}  \notag \\
& =\frac{1}{\left \vert \sqrt{1+N_{r_{1}}\left[ 1-\Upsilon ^{2}(\eta ,\beta
,\lambda _{1})\right] }\right \vert ^{2}},  \tag{B6}
\end{align}%
where we have used the Eq. (A4) and the integral formular 
\begin{align}
& \int \frac{d^{2}\gamma }{\pi }\exp (\varsigma \left \vert \gamma \right
\vert ^{2}+\xi \gamma +\omega \gamma ^{\ast }+f\gamma ^{2}+g\gamma ^{\ast 2})
\notag \\
& =\frac{1}{\sqrt{\varsigma ^{2}-4fg}}\exp \left( \frac{-\varsigma \xi
\omega +\xi ^{2}g+\omega ^{2}f}{\varsigma ^{2}-4fg}\right) .  \tag{B7}
\end{align}%
Likewise, by using the completeness relation of Fock states, one can get the
TMSVS $\left \vert \psi _{S}\left( r_{2}\right) \right \rangle $ can be
expanded in the basis of Fock states, i.e., 
\begin{equation}
\left \vert \psi _{S}\left( r_{2}\right) \right \rangle =\sec
hr_{2}\sum_{n=0}^{\infty }\left( -\tanh r_{2}\right) ^{n}\left \vert
n,n\right \rangle .  \tag{B8}
\end{equation}%
Then, utilizing the Eq. (B8), one can finally obtain the lower bound of the
fidelity for the TMSVS in the presence of the photon-loss environment 
\begin{align}
& F_{Q_{1}}\left( \beta \right) _{TMSV}  \notag \\
& =|\left \langle \psi _{S}\left( r_{2}\right) \right \vert \left[ \eta
e^{-i\beta }+\left( 1-\eta \right) e^{i\beta \lambda _{1}}\right] ^{\hat{n}}
\notag \\
& \sec hr_{2}\sum_{n=0}^{\infty }\left( -\tanh r_{2}\right) ^{n}\left \vert
n,n\right \rangle |^{2}  \notag \\
& =|\sec h^{2}r_{2}\sum_{n=0}^{\infty }\left[ \tanh ^{2}r_{2}\left[ \eta
e^{-i\beta }+\left( 1-\eta \right) e^{i\beta \lambda _{1}}\right] \right]
^{n}|^{2}  \notag \\
& =\frac{1}{\left \vert 1+\left. N_{r_{2}}\left[ 1-\Upsilon (\eta ,\beta
,\lambda _{1})\right] \right/ 2\right \vert ^{2}}.  \tag{B9}
\end{align}

\textbf{Appendix C: The Proof of Eq. (\ref{28})}

\bigskip Using the completeness relation of momentum states,%
\begin{equation}
\int_{-\infty }^{\infty }dp_{E^{\prime }}\left \vert p_{E^{\prime }}\right
\rangle \left \langle p_{E^{\prime }}\right \vert =1,  \tag{C1}
\end{equation}%
where $\left \vert p_{E^{\prime }}\right \rangle $ is the eigenstate of the
momentum operator $\hat{p}_{E^{\prime }}$ \cite{36,38} 
\begin{align}
& \left \vert p_{E^{\prime }}\right \rangle  \notag \\
& =\pi ^{-1/4}\exp \left( -\frac{1}{2}p_{E^{\prime }}^{2}+i\sqrt{2}%
p_{E^{\prime }}\hat{a}_{E^{\prime }}^{\dagger }+\frac{\hat{a}_{E^{\prime
}}^{\dagger 2}}{2}\right) \left \vert 0_{E^{\prime }}\right \rangle , 
\tag{C2}
\end{align}%
one can obtain%
\begin{align}
& \left \langle \Psi _{S+E^{\prime }}(x)|\Psi _{S+E^{\prime }}(x+\beta
)\right \rangle  \notag \\
& =\left \langle \Phi _{S+E^{\prime }}(x)\right \vert e^{\left. -ix\lambda
_{2}\hat{p}_{E^{\prime }}\right/ 2\kappa }e^{\left. i(x+\beta )\lambda _{2}%
\hat{p}_{E^{\prime }}\right/ 2\kappa }\left \vert \Phi _{S+E^{\prime
}}(x+\beta )\right \rangle  \notag \\
& =\left \langle \psi _{S}\right \vert e^{i\beta (\lambda _{2}-1)\hat{n}%
}\left \vert \psi _{S}\right \rangle \left \langle 0_{E^{\prime }}\right
\vert e^{\left. i\beta \lambda _{2}\hat{p}_{E^{\prime }}\right/ 2\kappa
}\left \vert 0_{E^{\prime }}\right \rangle  \notag \\
& =\left \langle \psi _{S}\right \vert e^{i\beta (\lambda _{2}-1)\hat{n}%
}\left \vert \psi _{S}\right \rangle \left \langle 0_{E^{\prime }}\right
\vert e^{\left. i\beta \lambda _{2}\hat{p}_{E^{\prime }}\right/ 2\kappa } 
\notag \\
& \times \int_{-\infty }^{\infty }dp_{E^{\prime }}\left \vert p_{E^{\prime
}}\right \rangle \left \langle p_{E^{\prime }}|0_{E^{\prime }}\right \rangle
\notag \\
& =e^{\left. -\beta ^{2}\lambda _{2}^{2}\right/ 16\kappa ^{2}}\left \langle
\psi _{S}\right \vert e^{i\beta (\lambda _{2}-1)\hat{n}}\left \vert \psi
_{S}\right \rangle ,  \tag{C3}
\end{align}%
where we have utilized the integrational formula%
\begin{equation}
\int_{-\infty }^{\infty }\exp (-hy^{2}+gy)dy=\sqrt{\pi /h}\exp (g^{2}/4h). 
\tag{C4}
\end{equation}%
Therefore, the lower bound of the fidelity in the phase diffusion
environment can be given by%
\begin{equation}
F_{Q_{2}}\left( \beta \right) =\Theta (\kappa ,\beta ,\lambda _{2})\left
\vert \left \langle \psi _{S}\right \vert e^{i\beta \left( \lambda
_{2}-1\right) \hat{n}}\left \vert \psi _{S}\right \rangle \right \vert ^{2}.
\tag{C5}
\end{equation}

\end{document}